\documentclass[twocolumn,twocolappendix]{aastex631}
\usepackage{graphicx,natbib,bm,url,color,amsmath}
\graphicspath{{./fig/}}
\newcommand{\EQ}{\begin{equation}}
\newcommand{\EN}{\end{equation}}
\newcommand{\EQA}{\begin{eqnarray}}
\newcommand{\ENA}{\end{eqnarray}}

\newcommand{\Eq}[1]{Equation~(\ref{#1})}
\newcommand{\Eqs}[2]{Equations~(\ref{#1}) and~(\ref{#2})}

\newcommand{\App}[1]{Appendix~\ref{#1}}

\newcommand{\Sec}[1]{Section~\ref{#1}}

\newcommand{\Fig}[1]{Figure~\ref{#1}}
\newcommand{\FFig}[1]{Figure~\ref{#1}}

\newcommand{\Tab}[1]{Table~\ref{#1}}

\newcommand{\bra}[1]{\langle #1\rangle}
\newcommand{\bbra}[1]{\left\langle #1\right\rangle}

{}
{}

{}
{}

{}
{}
{}
{}
{}
{}
{}
{}
{}
{}
{}
{}
{}
{}

{}

\newcommand{\hatkk}{\hat{\bm{k}}}

{}

{}
{}
{}

\newcommand{\kk}{\bm{k}}

\newcommand{\qq}{\bm{q}}
\newcommand{\xx}{\bm{x}}

\newcommand{\BB}{\bm{B}}

\newcommand{\oo}{\bm{\omega}}

\newcommand{\uu}{\bm{u}}

\newcommand{\SSS}{\bm{S}}

\newcommand{\nab}{{\bm{\nabla}}}
\newcommand{\GG}{\mbox{\boldmath $G$} {}}

\newcommand{\SSSS}{\mbox{\boldmath ${\sf S}$} {}}

\newcommand{\ii}{{\rm i}}

\newcommand{\DD}{{\rm D} {}}

\newcommand{\dd}{{\rm d} {}}

\def\ga{\mathrel{\mathchoice {\vcenter{\offinterlineskip\halign{\hfil
$\displaystyle##$\hfil\cr>\cr\sim\cr}}}
{\vcenter{\offinterlineskip\halign{\hfil$\textstyle##$\hfil\cr>\cr\sim\cr}}}
{\vcenter{\offinterlineskip\halign{\hfil$\scriptstyle##$\hfil\cr>\cr\sim\cr}}}
{\vcenter{\offinterlineskip\halign{\hfil$\scriptscriptstyle##$\hfil\cr>\cr\sim\cr}}}}}
\def\H{\mbox{\rm H}}

\def\Sp{\mbox{\rm Sp}}
\def\Ma{\mbox{\rm Ma}}

\def\Pm{\mbox{\rm Pr}_{\rm M}}

\def\Rey{\mbox{\rm Re}}

\def\EK{E_{\rm K}}

\def\csz{c_{\rm s0}}

\def\epsK{\epsilon_{\rm K}}

\def\urms{u_{\rm rms}}
\def\umax{u_{\rm max}}

\def\divurms{(\nab\cdot\uu)_{\rm rms}}
\def\divurms{d_{\rm rms}}
\def\drms{d_{\rm rms}}
\def\orms{\omega_{\rm rms}}

\newcommand{\W}{\,{\rm W}}
\newcommand{\V}{\,{\rm V}}

\newcommand{\T}{\,{\rm T}}
\newcommand{\G}{\,{\rm G}}

\newcommand{\K}{\,{\rm K}}
\newcommand{\g}{\,{\rm g}}
\newcommand{\s}{\,{\rm s}}

\newcommand{\m}{\,{\rm m}}

\newcommand{\J}{\,{\rm J}}

\newcommand{\A}{\,{\rm A}}

\def\apjl{ApJL}
\def\apjs{ApJS}

\def\aap{A\&A}

\begin{document}

\title{No evidence of vorticity production from irrotational turbulent gravitational collapse yet}

\author[0000-0002-7304-021X]{Axel Brandenburg}
\affiliation{Nordita, KTH Royal Institute of Technology and Stockholm University, Hannes Alfv\'ens v\"ag 12, SE-10691 Stockholm, Sweden}
\affiliation{The Oskar Klein Centre, Department of Astronomy, Stockholm University, AlbaNova, SE-10691 Stockholm, Sweden}
\affiliation{McWilliams Center for Cosmology \& Department of Physics, Carnegie Mellon University, Pittsburgh, PA 15213, USA}
\affiliation{School of Natural Sciences and Medicine, Ilia State University, 3-5 Cholokashvili Avenue, 0194 Tbilisi, Georgia}

\author[0000-0002-4324-0034]{Evangelia Ntormousi}
\affiliation{Scuola Normale Superiore, Piazza dei Cavalieri 7, 56126 Pisa, Italy}

\author[0000-0001-7888-6671]{Jennifer Schober}
\affiliation{Argelander-Institut f\"ur Astronomie, Universit\"at Bonn, Auf dem H\"ugel 71, 53121 Bonn, Germany}

\begin{abstract}
Gravitational collapse creates large amounts of kinetic energy that could potentially seed turbulence.
If such turbulence were also suitable to initiate dynamo action, the resulting magnetic field would further modify the dynamics, especially on small length scales.
However, a small-scale dynamo is believed to require vortical turbulence, whereas the collapse produces mainly irrotational motions, which may not be efficient for dynamo action.
Here, we study the efficiency of vorticity production during a turbulent collapse.
We adopt a barotropic equation of state, where pressure and density gradients are parallel,
and no magnetic field, so that vorticity can only be produced by viscosity
and by vortex dynamo-type amplification.
Using direct numerical simulations of gravitational collapse, we show that,
for the parameter space accessible to our numerical resolution,
these effects are related to the initial irrotational turbulence and are not induced by the collapse.
Vorticity production along with the associated small-scale dynamo action are still expected to occur for sufficiently large Reynolds numbers,
especially when baroclinic effects are allowed for,
but some of the earlier numerical evidence in the literature using a barotropic equation of state
is now found to be the result of subgrid-scale modeling and not reproduced in direct numerical simulations.
\end{abstract}

\section{Introduction}

It is generally believed that gravitational collapse generates turbulence \citep{field_2008, klessen_hennebelle_2010}.
While this is indeed quite plausible, it is unclear 
if numerical simulations have successfully demonstrated this conversion \citep{Hennebelle21,BN22,BN25} and what its efficiency would be.
This question is particularly important in the context of small-scale dynamo action driven by such a collapse \citep{Sur+10,Sur+12,XL20b,XL20},
because it could result in an amplification of magnetic energy below the Jeans scale \citep{Schober2026}, which would otherwise not occur \citep{BN22}.

Small-scale dynamos operate predominantly due to vortical turbulence; see \cite{BN25} for a recent discussion.
On the other hand, gravitational collapse produces predominantly irrotational flows \citep{Federrath+11b}.
Similar circumstances also apply to the cosmological dynamics of large-scale structure formation \citep{Vazza+17}.
The question of vorticity production from an irrotational collapse (in the absence of a magnetic field or driving forces) becomes clear-cut
when an isothermal or barotropic equation of state is employed, because then vorticity production via the baroclinic term would be impossible;
see, e.g., \cite{Vazquez-Semadeni+96}, \cite{DSB11}, \cite{Kapyla+18}, and \cite{Elias-Lopez+23, Elias-Lopez+24} for earlier studies,
where different sources of vorticity were identified (baroclinicity, shear, and rotation).
In realistic flows, baroclinicity could be provided by cooling \citep{VS22}.
However, in the present context, the only way to produce vorticity from purely irrotational flows is via the interaction with viscosity.

Earlier work by \cite{BN22,BN25} has already addressed the question of turbulence and small-scale dynamo action from the collapse
and concluded that the collapse can enhance the magnetic field by compression, but this alone is not a dynamo.
The same work showed that a dynamo did occur in a collapsing flow setup, but it was powered by the decay of the initial turbulence rather than the collapse itself.
Specifically, dynamo action occurred as long as the magnetic Reynolds number was still supercritical for dynamo action.
Distinguishing between these two mechanisms, i.e., collapse-generated versus decaying turbulence under the influence of compression,
was facilitated by transforming the equations into a collapsing coordinate system
\citep{BN25}, which was based on a formalism first proposed by \cite{Irshad+26}.

An important step toward a demonstration of collapse-induced turbulence
was based on a decomposition into radial and transverse velocity components \citep{Hennebelle21}.
This related to earlier work on turbulence amplification during a collapse \citep{Robertson+Goldreich12,Murray+17,VS20}.
However, none of these works focused on vorticity production, which is of particular interest for dynamo action.
In the aforementioned papers on collapse-induced small-scale dynamo action \citep{Sur+10,Sur+12,XL20b,XL20},
the evidence for dynamo action was based on an excess over the purely compression-related amplification.
They employed a slightly supercritical Bonnor--Ebert sphere \citep{Ebert55, Bonnor56} as the initial condition.
In a recent paper, also using a Bonnor--Ebert setup, \cite{Schober2026} found indications for small-scale dynamo action based on
a comparison between the measured and theoretically predicted growth rates of the rms magnetic field.

Here, we focus on the basic question of vorticity production.
To address this question, we employ the model of \cite{Schober2026} 
along with just irrotational velocity perturbations initially.
No magnetic fields are included, because they would only obscure our basic question regarding vorticity generation; see \cite{BS25}
for the identification of two distinct mechanisms for producing vorticity from magnetic fields.
Specifically, we reproduce here the setup of \cite{Schober2026}, but removed magnetic fields and the shock viscosity and used instead a constant kinematic viscosity.
We vary both the resolution and the value of the viscosity to determine what we consider to be numerically reliable limits.

\section{Our model}

\subsection{Basic equations}

Following \cite{Schober2026}, our computational domain is a periodic cube of size $L^3=(4\pi)^3$
and an isentropic equation of state with a ratio of specific heats being $\gamma=5/3$ is employed,
where the specific enthalpy $h$ is related to the density $\rho$ through \citep{MB06}
\begin{equation}
h=\frac{\csz^2}{\gamma-1}\left(\frac{\rho}{\rho_0}\right)^{\gamma-1}.
\end{equation}
Here, $\rho_0$ is the mean density, which is constant in time because of mass conservation, and $\csz$ is the sound speed when $\rho=\rho_0$.
The isentropic equation of state is similar to the isothermal one, which emerges in the limit $\gamma\to1$, when $h=\csz^2\ln(\rho/\rho_0)$.
The lowest wavenumber in the domain is $k_1=2\pi/L=0.5$.
A modified isothermal Bonnor--Ebert sphere of radius $R=2\pi$ is inserted at $|\xx|=0$, i.e., at the center of our domain.
For radii larger than $R$, the density is set equal to the value at $|\xx|=R$.
Owing to the use of periodic boundary conditions, the sphere repeats itself to infinity.

We stress that we perform direct numerical simulations, so no subgrid-scale modeling is done.
In particular, no shock viscosity is used because it makes the results harder to interpret.
Studying potential artifacts of specific subgrid-scale schemes would be interesting in its own right, but is not the purpose of the present study.
For a monatomic gas, the stress tensor is given by $2\rho\nu\SSSS$, where $\SSSS$ is the traceless rate-of-strain tensor with the components
$\mathsf{S}_{ij}=(\partial_i u_j+\partial_j u_i)/2-\delta_{ij}\nab\cdot\uu/3$, and $\uu$ is the velocity.
Our equations are \citep{Passot+95}
\begin{equation}
\nabla^2\Phi=4\pi G_\mathrm{N}\left(\rho-\rho_0\right),
\label{del2Phi}
\end{equation}
\begin{equation}
{\DD\uu\over\DD t}=-\nab\left(h+\Phi\right)+\frac{1}{\rho}\nab\cdot(2\rho\nu\SSSS),
\label{DuDt}
\end{equation}
\begin{equation}
{\DD\ln\rho\over\DD t}=-\nab\cdot\uu,
\label{Dlnrho}
\end{equation}
where $\Phi$ is the gravitational potential, $G_\mathrm{N}$ is Newton's constant, 
$\DD/\DD t=\partial/\partial t+\uu\cdot\nab$ is the advective derivative,
and $\nu$ is the kinematic viscosity, which is here assumed to be constant.

Here, we compute the kinetic energy dissipation as $\epsK=\bra{2\rho\nu\SSSS^2}$.
This also determines the Kolmogorov wavenumber $k_\nu=(\epsK/\rho_0\nu^3)^{1/4}$.
The vorticity is given by $\oo=\nab\times\uu$, and $\bra{\oo^2}/2$ is the enstrophy.
The rms values of $\oo$ and $\nab\cdot\uu$ are denoted as
\begin{equation}
\omega_\mathrm{rms}=\bra{(\nab\times\uu)^2}^{1/2},\quad
d_\mathrm{rms}=\bra{(\nab\cdot\uu)^2}^{1/2}.
\end{equation}
Note also that $\bra{2\SSSS^2}=\omega_\mathrm{rms}^2+(4/3)\,d_\mathrm{rms}^2$.
In our nearly irrotational simulations, its value is always dominated by the second term.

\subsection{Numerical aspects and initial condition} \label{sec:Summary_runs}

We use the {\sc Pencil Code} \citep{PC}, which employs
sixth-order centered differences and a third-order time-stepping scheme.
In four cases, we use fifth-order upwinding for the advective derivatives in \Eqs{DuDt}{Dlnrho}, i.e., we write \citep{Dobler+06}
\begin{equation}
-(\uu\cdot\nab)_\mathrm{upwind}=-\uu\cdot\nab+\nu_\mathrm{hyp}\nabla^6,
\end{equation}
where the first- and sixth-order derivatives on the rhs are sixth-order centered and second-order centered, respectively,
and $\nu_\mathrm{hyp}=|\uu|\,\delta x^5/60$ is a velocity-dependent hyperviscosity.
An upwinding derivative means that one uses one more data point of the advected variable in the upwind direction of the velocity
(here, three in the upwind direction and two in the downstream direction for our fifth-order scheme).
We adopt numerical resolutions between $512^3$ and $2048^3$ mesh points.

Our initial velocity is constructed in Fourier space as
$\uu(\xx)=\sum\tilde{\uu}(\kk)\,e^{\ii\kk\cdot\xx}$, where 
$\tilde{\uu}(\kk)$ is an irrotational velocity field given by
\begin{equation}
\tilde{u}_i(\kk)=u_\mathrm{ini}\,\hat{k}_i\hat{k}_j\,\tilde{S}_j(\kk).
\end{equation}
Here, $u_\mathrm{ini}$ is an amplitude factor, $\hat{k}_i$ is the $i$th component
of the unit vector $\hatkk\equiv\kk/k$, $\tilde{S}_j(\kk)$
is a vector field in Fourier space with three independent components $j$
that depend on $k=|\kk|$, and have random phases $\varphi(\kk)$ for each $\kk$ vector.
Here, we choose
\begin{equation}
\tilde{S}_j(\kk)=\frac{k_0^{-3/2} (k/k_0)^{\alpha/2-1}}
{1+(k/k_0)^{(\alpha+5/3)/2}}\,e^{\left[\ii\varphi(\kk)-\kk^2/k_\mathrm{cut}^2\right]},
\label{kcut}
\end{equation}
where $k_0$ is the peak wavenumber of the initial condition and $\alpha$ is the slope of the subinertial range.

The isothermal Bonnor--Ebert sphere is a solution to the Emden--Chandrasekhar equation,
which is the isothermal version of the Lane--Emden equation, i.e.,
\begin{equation}
\frac{1}{r^2}\frac{\dd}{\dd r}\left(r^2\frac{\dd\psi}{\dd r}\right)=e^{-\psi}
\end{equation}
with the boundary conditions $\psi=1$ and $\dd\psi/\dd r=0$ at $r=0$; see the reviews of \cite{Larson03} and \cite{McKee+Ostriker07},
and the early work by \cite{Larson69}, \cite{Penston69}, and \cite{Shu77} on gravitational collapse computations.
The density is usually given by $\rho=\exp(-\psi)$, but \cite{Schober2026}
used\footnote{\tiny\url{https://github.com/JenSchober/publications/tree/master/2026/SchoberEtAl_AA/}} $\rho=\exp[\exp(-\psi)]$.
This implies that $\rho_0\equiv\bra{\rho}\approx1.15$ instead of $\approx0.133$.
To facilitate comparison with their work, we decided to adopt their initial density profile,
but see \App{DifferentDensityProfiles} for a comparison of the evolution in both cases.

\begin{table*}[htb]\caption{
Summary of our simulations for different values of $u_\mathrm{ini}$ and $\nu$.
For $t=2.7$, we give the values of $k_\nu$, $\epsK$, $\divurms$, $\orms$, as well as the mesh and actual Reynolds numbers.
We further list the early peak values of the terms $T_\mathrm{gen0}$, $T_\mathrm{dis0}$, and $T_\mathrm{dyn0}$,
as well as the ratios $R_\mathrm{gen1}$, $R_\mathrm{gen2}$, and $R_\mathrm{dyn1}$.
Dashes indicate that, due to the lack of a local plateau ($R_\mathrm{gen1}$) or a local maximum ($R_\mathrm{gen2}$), no values could be determined.
The last column gives the number of mesh points per direction, and an asterisk denotes runs with fifth order upwinding.
}\hspace{-18mm}\vspace{12pt}\centerline{\begin{tabular}{cccc cccc cccc cccc c}
Run & $u_\mathrm{ini}$ & $k_0$ & $\nu$ & $k_\nu$ & $\epsK$ & $\!\!\divurms\!\!$ & $\orms$ & $\!\!\Rey_\mathrm{m}\!\!$ & $\Rey$ &
$T_\mathrm{dyn0}$ & $T_\mathrm{gen0}$ & $T_\mathrm{dis0}$ & $R_\mathrm{gen1}$ & $R_\mathrm{gen2}$ & $R_\mathrm{dyn1}$ & $N$ \\
\hline
A & $0.20$ & $ 2.5$ & $0.005$ & $24.2$ & $0.050$ & $1.90$ & $0.220$ & $4.1$ & $ 80$ & $0.2700$ & $0.970$ & $1.200$ & $0.63$ &   ---  & $0.190$ & $2048*$ \\% Hy2048b_noshock2
B & $0.20$ & $ 2.5$ & $0.010$ & $14.4$ & $0.049$ & $1.40$ & $0.130$ & $2.0$ & $ 40$ & $0.0430$ & $0.150$ & $0.170$ & $0.69$ & $1.13$ & $0.220$ & $2048$ \\% Hy2048a_noshock
C & $0.20$ & $ 2.5$ & $0.010$ & $14.0$ & $0.044$ & $1.30$ & $0.120$ & $3.7$ & $ 40$ & $0.0330$ & $0.140$ & $0.160$ & $0.64$ & $1.09$ & $0.170$ & $1024$ \\% Hy1024b_noshock
D & $0.20$ & $ 2.5$ & $0.014$ & $10.9$ & $0.044$ & $1.20$ & $0.085$ & $2.6$ & $ 28$ & $0.0110$ & $0.088$ & $0.064$ & $0.69$ & $1.02$ & $0.160$ & $1024$ \\% Hy1024c_noshock
E & $0.20$ & $ 2.5$ & $0.020$ & $ 8.4$ & $0.045$ & $1.00$ & $0.061$ & $1.8$ & $ 20$ & $0.0034$ & $0.120$ & $0.086$ & $0.66$ & $0.93$ & $0.120$ & $1024$ \\% Hy1024a_noshock
F & $0.20$ & $ 2.5$ & $0.020$ & $ 8.4$ & $0.045$ & $1.00$ & $0.069$ & $3.5$ & $ 20$ & $0.0059$ & $0.120$ & $0.088$ & $0.63$ & $0.97$ & $0.140$ & $ 512$ \\% Hy512d_noshock
G & $0.20$ & $ 2.5$ & $0.030$ & $ 6.2$ & $0.047$ & $0.87$ & $0.048$ & $2.2$ & $ 13$ & $0.0024$ & $0.160$ & $0.120$ & $0.61$ & $0.88$ & $0.096$ & $ 512$ \\% Hy512e_noshock
H & $0.20$ & $ 2.5$ & $0.040$ & $ 5.1$ & $0.051$ & $0.80$ & $0.034$ & $0.9$ & $ 10$ & $0.0019$ & $0.170$ & $0.130$ & $0.60$ & $0.82$ & $0.062$ & $1024$ \\% Hy1024d_noshock
I & $0.20$ & $ 2.5$ & $0.040$ & $ 5.1$ & $0.050$ & $0.79$ & $0.038$ & $1.7$ & $ 10$ & $0.0020$ & $0.180$ & $0.140$ & $0.60$ & $0.84$ & $0.070$ & $ 512$ \\% Hy512g_noshock
J & $0.10$ & $ 2.5$ & $0.020$ & $ 7.3$ & $0.027$ & $0.79$ & $0.013$ & $3.5$ & $ 20$ & $0.0000$ & $0.007$ & $0.005$ & $0.73$ &   ---  & $0.036$ & $ 512$ \\% Hy512d_noshock_ampl01
K & $0.05$ & $ 2.5$ & $0.020$ & $ 6.8$ & $0.020$ & $0.71$ & $0.004$ & $3.4$ & $ 20$ & $0.0000$ & $0.000$ & $0.000$ & $0.74$ &   ---  & $0.011$ & $ 512$ \\% Hy512d_noshock_ampl005
L & $0.20$ & $ 5.0$ & $0.020$ & $ 7.3$ & $0.027$ & $0.80$ & $0.032$ & $3.4$ & $ 10$ & $0.0046$ & $0.230$ & $0.170$ &   ---  &   ---  & $0.066$ & $ 512$ \\% Hy512d_noshock_k10
M & $0.20$ & $10.0$ & $0.020$ & $ 6.7$ & $0.019$ & $0.70$ & $0.011$ & $3.3$ & $  5$ & $0.0033$ & $0.260$ & $0.210$ &   ---  &   ---  & $0.024$ & $ 512$ \\% Hy512d_noshock_k20
N & $0.20$ & $ 2.5$ & $0.010$ & $13.9$ & $0.043$ & $1.40$ & $0.130$ & $7.2$ & $ 40$ & $0.0490$ & $0.150$ & $0.700$ & $0.12$ &   ---  & $0.054$ & $ 512*$ \\% Hy512h_noshock2
O & $0.20$ & $ 5.0$ & $0.010$ & $12.0$ & $0.024$ & $1.00$ & $0.062$ & $6.9$ & $ 20$ & $0.0180$ & $0.160$ & $0.370$ &   ---  &   ---  & $0.046$ & $ 512*$ \\% Hy512h_noshock_k10
P & $0.20$ & $10.0$ & $0.010$ & $10.1$ & $0.012$ & $0.78$ & $0.027$ & $6.7$ & $ 10$ & $0.0150$ & $0.240$ & $0.270$ &   ---  &   ---  & $0.050$ & $ 512*$ \\% Hy512h_noshock_k20
Q & $0.20$ & $ 2.5$ & $0.020$ & $ 6.6$ & $0.002$ & $0.75$ & $0.180$ & $1.1$ & $ 20$ & $0.0440$ & $0.440$ & $0.370$ & $0.41$ & $0.92$ & $0.220$ & $ 512$ \\% Hy512d_noshock_nopretend2
\label{TSummary}\end{tabular}}\end{table*}

Throughout this paper, we employ nondimensional units by setting
\begin{equation}
\csz=2k_1=\rho_0/1.15=1.
\end{equation}
This implies that velocities are measured in units of $\csz$,
lengths in units of $(2\,k_1)^{-1}$, time in units of $(2\,k_1 \csz)^{-1}$, and density in units of $\rho_0/1.15$.
With the exception of \Sec{NoncollapsingTurbulence}, where we compare with noncollapsing turbulence ($G_\mathrm{N}=0$), we set $4\pi G_\mathrm{N}=1$.

In the following, we mostly use $k_0=2.5$, $k_\mathrm{cut}=25$, and $\alpha=4$.
The setup with these parameters agrees with those of Runs~H1b and H2b of \cite{Schober2026}, except that our initial velocity is irrotational
and we have also used a finite value of $k_\mathrm{cut}$.
In fact, we have chosen $k_\mathrm{cut}$ to be much smaller than the Nyquist wavenumber, $k_\mathrm{Ny}=\pi/\delta x$,
where $\delta x$ is the mesh spacing.
For our numerical resolutions between $512^3$ and $2048^3$ mesh points, we have $k_\mathrm{Ny}=128$ and 512, respectively.
The reason for our small value of $k_\mathrm{cut}$ is that we wanted to make sure no vorticity is initially present at the mesh scale.

A useful measure of the value of the viscosity is the mesh Reynolds number, $\Rey_\mathrm{m}=\umax \delta x/\nu$.
Here, $\delta x=L/N$ with $L=4\pi$ being the size of the domain.
Earlier experience \citep{Bra03} shows that for smooth, nearly incompressible flows, $\Rey_\mathrm{m}$ can well be as large as 50,
but for compressive flows, it might need to be well below a value of around 0.5.
This is especially clear when shocks form.
Our usual Reynolds number is defined as $\Rey=\urms/(\nu k_0)$.
We also present velocity, logarithmic density, and vorticity spectra, $\Sp(\uu)$, $\Sp(\ln\rho)$, and $\Sp(\oo)$, respectively.
They are normalized such that $\int\Sp(\uu)\,\dd k=\bra{\uu^2}/2$, $\int\Sp(\ln\rho)\,\dd k=\bra{(\ln\rho)^2}/2$, and $\int\Sp(\oo)\,\dd k=\bra{\oo^2}/2$.
Note that $\rho_0\,\Sp(\uu)$ is the usual kinetic energy spectrum, which is also often called $\EK(k)$.
The spectrum of the irrotational part of the velocity, $\uu_\mathrm{irro}$ is given by $\Sp(\uu_\mathrm{irro})=\Sp(\uu)-\Sp(\oo)/k^2$.

\subsection{Vorticity production} \label{sec:Vorticity-Production}

To determine the terms leading to vorticity production, we take the curl of \Eq{DuDt} and obtain
\begin{equation}
\frac{\partial\oo}{\partial t}=\nab\times\left[\uu\times\oo+\nu\left(\GG-\nab\times\oo\right)\right],
\label{dodt}
\end{equation}
where $G_i=2{\sf S}_{ij}\nabla_j\ln\rho$ is the crucial vector whose curl drives vorticity.
The other two terms vanish initially, when there is no vorticity \citep{MB06}.
Defining $\qq=\nab\times\oo$, and taking the dot product of \Eq{dodt} with $\oo$,
we obtain after volume averaging (denoted by angle brackets) and integration by parts
\begin{equation}
\frac{\dd}{\dd t}\bbra{\dfrac{\oo^2}{2}}
=\bra{\qq\cdot(\uu\times\oo)}
+\nu\left(\bra{\qq\cdot\GG}-\bra{\qq^2}\right).
\label{vortprod}
\end{equation}
When upwinding is used, additional terms of the order of $\delta x^5$ appear.

It should be noted that, even in the absence of viscosity, vorticity can be produced by curved shocks \citep{Truesdell52,Lighthill57,Hayes57}.
For applications to the interstellar medium, see the work by \cite{Fleck+91}.
This is because shocks are strictly discontinuous and not isentropic.
In fact, \Eq{vortprod} only applies to differentiable flows, which is also what is assumed when using the {\sc Pencil Code}.

Except for the term involving $\GG$, this equation is analogous to the induction equation,
where $\bra{\qq\cdot(\uu\times\oo)}$ is the term analogous to the magnetic energy generation by the induction term
and $\nu\bra{\qq^2}$ is analogous to the magnetic energy loss by the dissipation term.
In this work, we monitor all three terms, along with the time derivative of the enstrophy, $\dd\bra{\oo^2/2}/\dd t$.
Here, our use of volume averages precludes the investigation of enstrophy fluxes within the domain.
Those would be of interest if one wanted to quantify the local vorticity accumulation and amplification caused by the contraction, for example.
However, we do briefly return to this point later in the paper.

\section{Results}

\subsection{Numerical constraints on the viscosity}

\begin{figure}\begin{center}
\includegraphics[width=\columnwidth]{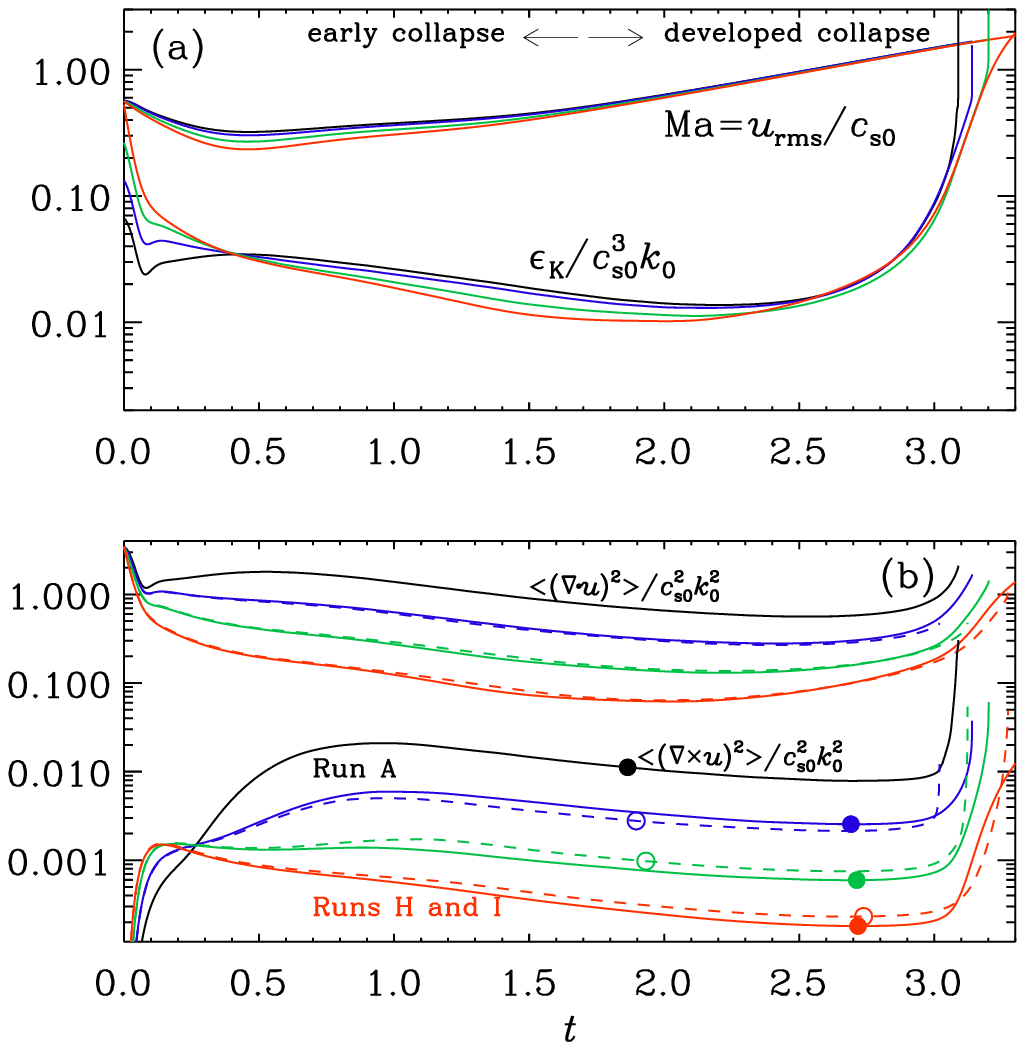}
\end{center}\caption{
(a) Time series of the Mach number, $\Ma=\urms/\csz$, for different values of $\nu$, along with
the nondimensional kinetic energy dissipation, $\epsK/\csz^3 k_0$, as well as the time dependence
of (b) $\bra{(\nab\cdot\uu)^2}/\csz^2 k_0^2$ and $\bra{(\nab\times\uu)^2}/\csz^2 k_0^2$, for
Runs~A (black), B (blue), E (green), and H (red).
The corresponding dashed lines denote Runs~C, F, and I, which are lower-resolution versions of Runs~B, E, and H, respectively.
The filled and open symbols on the solid and dashed lines, respectively,
indicate the times when the mesh Reynolds number exceeds the value 2,
after which the runs may no longer be fully reliable.
In (a), the arrows indicate the early and developed collapse phases of the runs.
}\label{pcomp_ts}\end{figure}

We recall that we perform direct numerical simulations, where dissipation
is exclusively accomplished by the explicit viscosity, as quantified by the value of $\nu$; see \Tab{TSummary} for a summary of our runs.
We begin with a survey for different values of $\nu$ and also different numbers of mesh points, $N^3$.
In \Fig{pcomp_ts}, we show the time evolution of the Mach number, $\Ma=\urms/\csz$, for different values of $\nu$.
We see that the growth rate of the velocity changes at $t\approx1.7$.
To distinguish between the early and late phases, we decided to talk here about
``early collapse'' and ``developed collapse'', as indicated in \Fig{pcomp_ts}.
In all cases, the collapse ends with a finite-time singularity.
The time when this happens is called the collapse time, which is of the order of the initial free-fall time.
We also show the nondimensional kinetic energy dissipation, $\epsK/\csz^3 k_0$, as well as the time dependence
of $\bra{(\nab\cdot\uu)^2}/\csz^2 k_0^2$ and $\bra{(\nab\times\uu)^2}/\csz^2 k_0^2$.

%FIG1

We see that the collapse occurs at a time between $t=3$ and $3.2$ with the lowest value of $\nu$ being the case with the earliest collapse.
Until this time, all time traces are broadly similar, except that the more viscous runs display an early phase of more dissipation.
This can be seen in the slightly lower values of the Mach number at $t=0.5$ for the more viscous runs and the slightly larger dissipation before that time.
The rms velocity starts to grow exponentially at the expected growth rate, which is insensitive to the value of $\nu$.
In the early phase of the exponential growth, the energy dissipation is less for the less viscous runs, but it becomes
larger during the main phase of the collapse after $t=2.5$.
After the time $t\approx2.7$, the runs may no longer be reliable at the present resolution.
The exact time, however, depends on the run.

\begin{figure}\begin{center}
\includegraphics[width=\columnwidth]{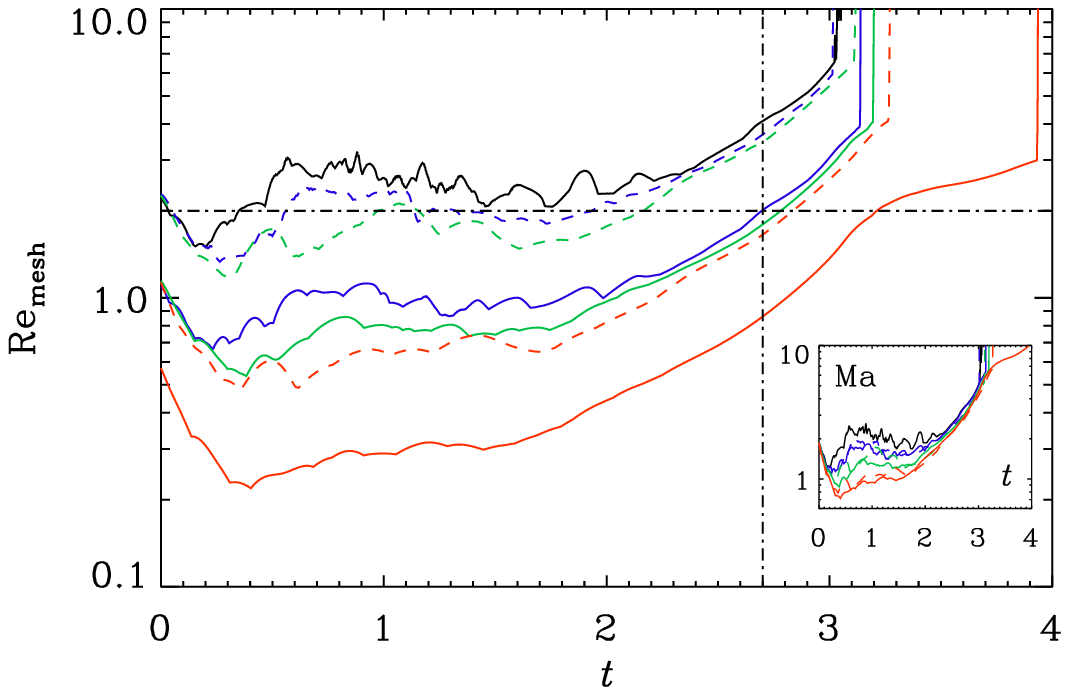}
\end{center}\caption{
Mesh Reynolds number for the same runs as in Figure~\ref{pcomp_ts}, shown in the same colors and line styles.
The vertical dashed-dotted line marks the time $t=2.7$ beyond which we deem the runs no longer trustworthy,
while the horizontal dashed-dotted line marks $\Rey_\mathrm{m}=0.7$, above which we also deem the runs no longer trustworthy.
The inset shows the evolution of the Mach number based on the maximum velocity.
}\label{pcomp_mesh_Re}\end{figure}

In \Sec{sec:Summary_runs}, we discussed typical values of the mesh Reynolds number.
In the present case, we see that at $t=2.7$, the time we deemed to be the limit beyond which we can trust the results,
$\Rey_\mathrm{m}$ is between 2 and 4.
Conversely, if we postulate the limit to be given by $\Rey_\mathrm{m}=2$, we would expect the
$1024^3$ run with $\nu=0.02$ to be reliable until $t=2.7$, and the run with $1024^3$ run with $\nu=0.01$
would only be reliable until $t=1.9$; see \Fig{pcomp_mesh_Re}.
We point out that the trend with resolution is opposite for $\nu=0.01$ and 0.02:
for $\nu=0.01$, larger resolution leads to larger values of $\epsK$, $\divurms$, and $\orms$, while
for $\nu=0.02$, larger resolution leads to smaller values of $\epsK$, $\divurms$, and $\orms$.

%FIG2

%FIG3
\begin{figure*}\begin{center}
\includegraphics[width=\textwidth]{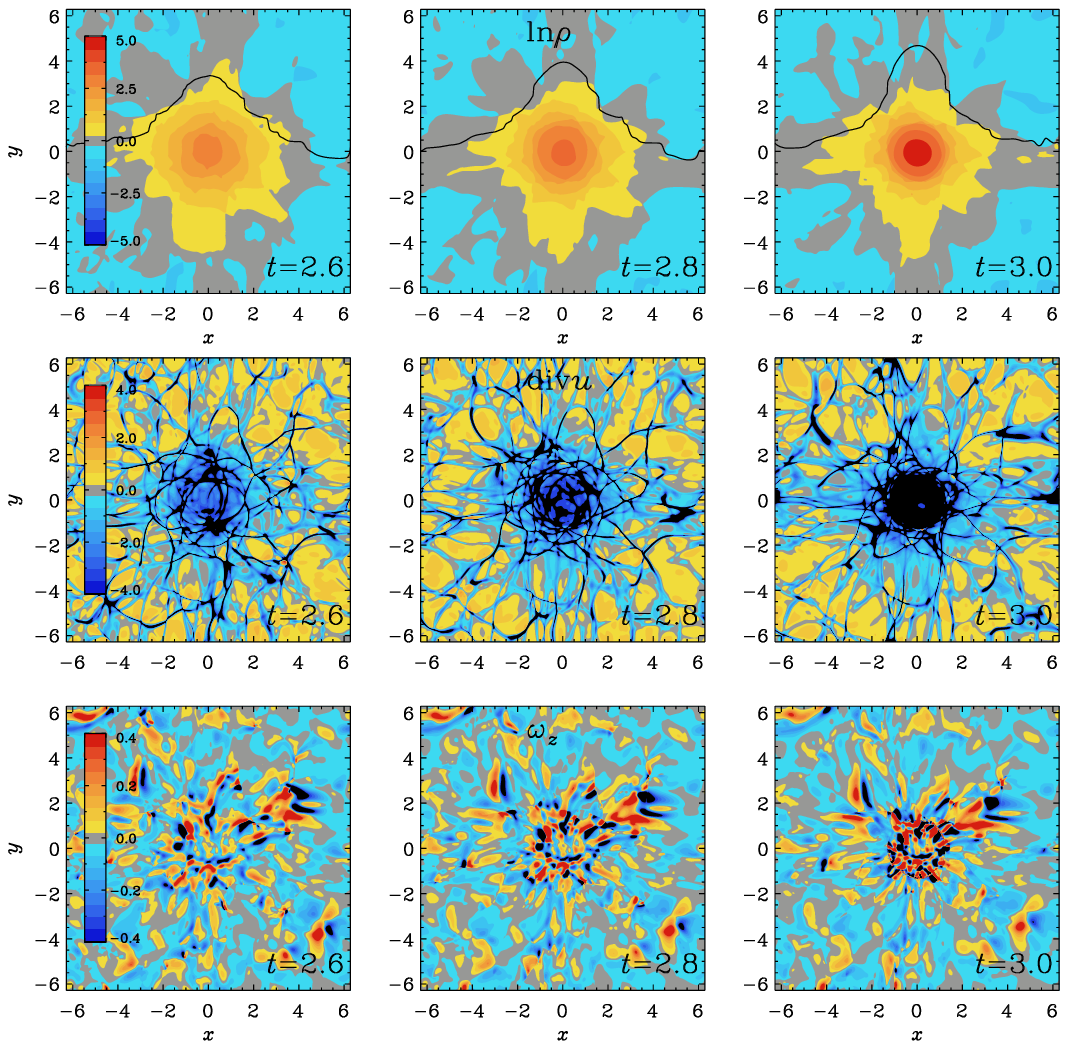}
\end{center}\caption{
Visualizations of $xy$ slices of $\ln\rho$ (top), $\nab\cdot\uu$ (middle) and $\omega_z$ (bottom) through $z=0$ at $t=2.6$, 2.8, and 3.0 (from left to right) for Run~A.
In the top row, the black line shows the profile of $\ln\rho(x,0)$ through the middle.
}\label{pVAR}\end{figure*}

Upwinding allows us to double the values of $\Rey$ and $\Rey_\mathrm{m}$ at the same numerical resolution; compare Runs~A and B.
On the other hand, runs with the same value of $\Rey$ (compare Run~N with Run~B) show that $\orms$ (and also $\drms$) are well converged,
but $T_\mathrm{dis0}$, and therefore also the ratios $R_\mathrm{gen1}$ and $R_\mathrm{dyn1}$, are not reliable when upwinding is used.
This demonstrates the high sensitivity of these quantities to numerical aspects.
Given that $\orms$ (and also $\drms$) are well converged, we can say that the problem lies in the diagnostics rather than the actual simulation.
Therefore, the fact that with upwinding, the nominal values of $T_\mathrm{dis}=\nu\bra{\qq^2}$ are so large suggests that the calculation of
the second derivatives in $\qq=\nab\times\nab\times\uu$ with a small number of mesh points becomes problematic.

At times before $t\approx1.7$, i.e., during the early phase of the collapse,
both $\bra{(\nab\cdot\uu)^2}\equiv d_\mathrm{rms}^2$ and $\bra{(\nab\times\uu)^2}\equiv\orms^2$ decay.
Only the less viscous runs display a subsequent growth of $d_\mathrm{rms}^2$ after the time $t=2$,
while $\orms^2$ continues to decay all the way until $t\approx2.7$, which is the time until we deem the runs still reliable.
For some of the runs with lower mesh Reynolds number, this point may be reached somewhat earlier.
One reason why some of the runs may become unreliable fairly early on is the fact that during the time interval $0.5\leq t\leq2$,
$\Rey_\mathrm{m}$ is nearly constant and thus mostly affected by the initial turbulence rather than the subsequent collapse;
see \Fig{pcomp_mesh_Re}.
Furthermore, already from the beginning of these runs, the Mach number based on the maximum velocity is mostly supersonic,
as shown in the inset of \Fig{pcomp_mesh_Re}.

Central $xy$ cross sections of $\ln\rho$, $\nab\cdot\uu$, and $\omega_z$ through $z=0$ at three times near the end of Run~A show that shocks do form.
However, they are not only in the central part of the domain, but also elsewhere; see \Fig{pVAR}.
This is due to the initially irrotational turbulence.
Those shocks can contribute to producing vorticity through viscosity.
Indeed, the images of $\omega_z$ in \Fig{pVAR} suggest that preexisting vorticity just gets more concentrated toward the center as the collapse proceeds.
The variations in $\ln\rho$ are much smoother and nearly flat in the middle.
This is demonstrated more clearly through the overlaid central line profile.

\subsection{Vorticity production terms}

We now analyze the three contributions to $\dd\bra{\oo^2/2}/\dd t$ in \Eq{vortprod}:
\begin{equation}
T_\mathrm{dyn}=\bra{\qq\cdot(\uu\times\oo)},\;
T_\mathrm{gen}=\nu\bra{\qq\cdot\GG},\;
T_\mathrm{dis}=\nu\bra{\qq^2}.
\end{equation}
The combination of all three contributions matches $\dd\bra{\oo^2/2}/\dd t$ during the time when the run is reliable, i.e.,
\begin{equation}
\dfrac{\dd}{\dd t}\bbra{\dfrac{\oo^2}{2}}
=T_\mathrm{dyn}+T_\mathrm{gen}-T_\mathrm{dis}.
\label{Tdyn_appearance}
\end{equation}
We also define the ratios $R_\mathrm{dyn}=T_\mathrm{dyn}/T_\mathrm{dis}$ and $R_\mathrm{gen}=T_\mathrm{gen}/T_\mathrm{dis}$ as functions of time.

\begin{figure}\begin{center}
\includegraphics[width=\columnwidth]{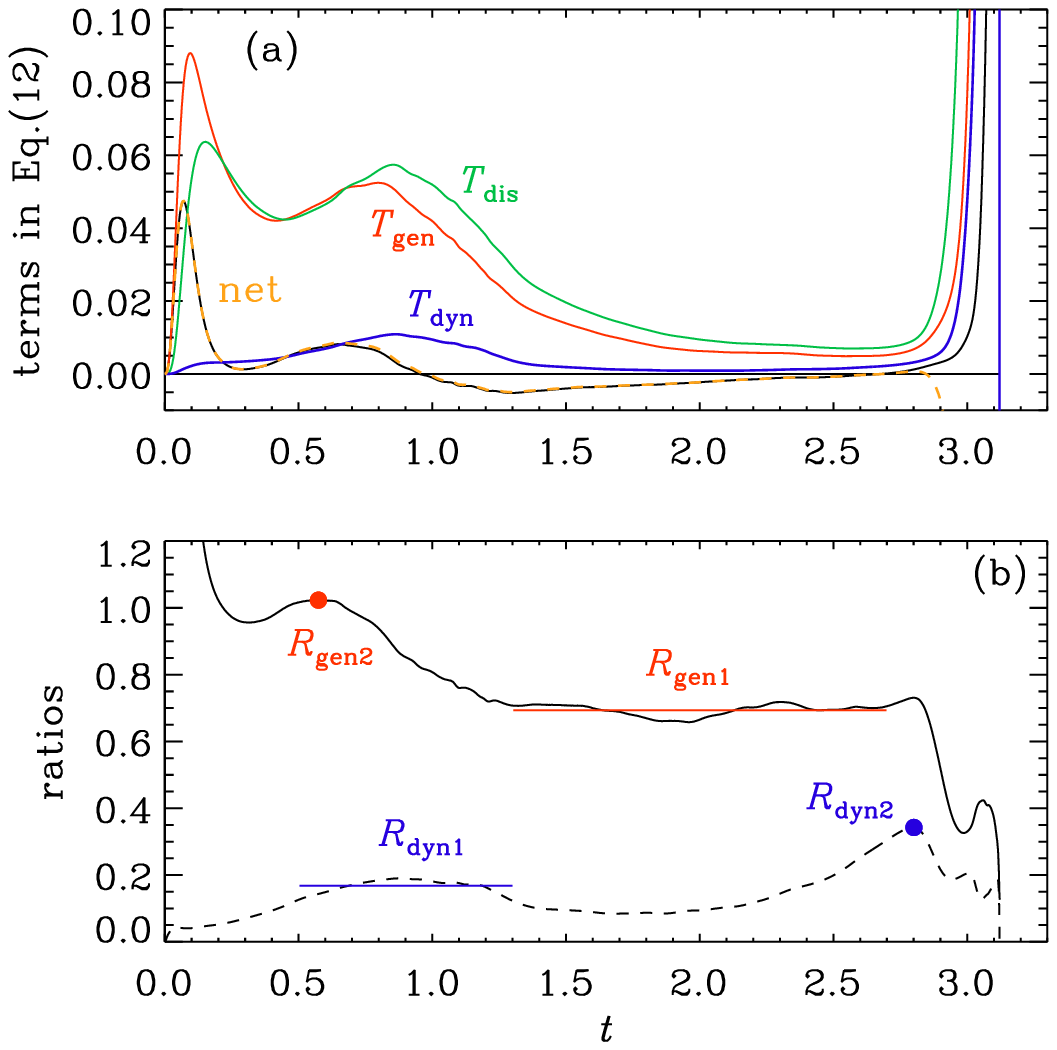}
\end{center}\caption{
(a) $\bra{\qq\cdot(\uu\times\oo)}$ (blue), $\nu\bra{\qq\cdot\GG}$ (red), and $\nu\bra{\qq^2}$ (green) for Run~D.
The dashed orange line represents the combination of these three terms and agrees with the
black line, which denotes the computed value of $\dd\bra{\oo^2/2}/\dd t$.
(b) ratios $\bra{\qq\cdot\GG}/\bra{\qq^2}$ (solid line) and $\bra{\qq\cdot(\uu\times\oo)}/\nu\bra{\qq^2}$ (dashed line).
The red filled symbol denotes the maximum of the first ratio after the initial transient before $t=0.2$,
the red line is the average of this ratio during the subsequent time interval from 1.3 to 2.7,
the blue line marks the average of the second ratio during the time interval from 0.5 to 1.3,
and the blue filled symbol marks the last maximum near the end of the run.
}\label{pom2_Hy1024c_noshock}\end{figure}

Each run is characterized by a very early adjustment phase ($t<0.2$) when vorticity is being produced by
the small departures from what should have been a perfectly irrotational flow.
At the same time when vorticity is being produced, 
also $T_\mathrm{dis}$ begins to grow and almost tracks $T_\mathrm{gen}$.
The ratio $T_\mathrm{gen}/T_\mathrm{dis} =\bra{\qq\cdot\GG}/\bra{\qq^2}$ 
drops and reaches a second (but lower) maximum during the second early phase ($0.5\leq t\leq0.7$),
and finally settles at an approximately constant value when this ratio is around 0.6; see \Tab{TSummary} for an overview of these values for different runs.
The dynamo-like term $T_\mathrm{dyn}$ is always subdominant, and the ratio $T_\mathrm{dyn}/T_\mathrm{dis}$ 
barely reaches 20\%; see \Fig{pom2_Hy1024c_noshock}.
It could in principle become dominant if the trend continues to lower viscosities, but that regime is not currently accessible numerically.
However, this is not the only trend.
As we will show in the following, there is also a dependence of the vorticity on the initial velocity.

%FIG4

\subsection{Viscosity dependence of vorticity generation}

\begin{figure}\begin{center}
\includegraphics[width=\columnwidth]{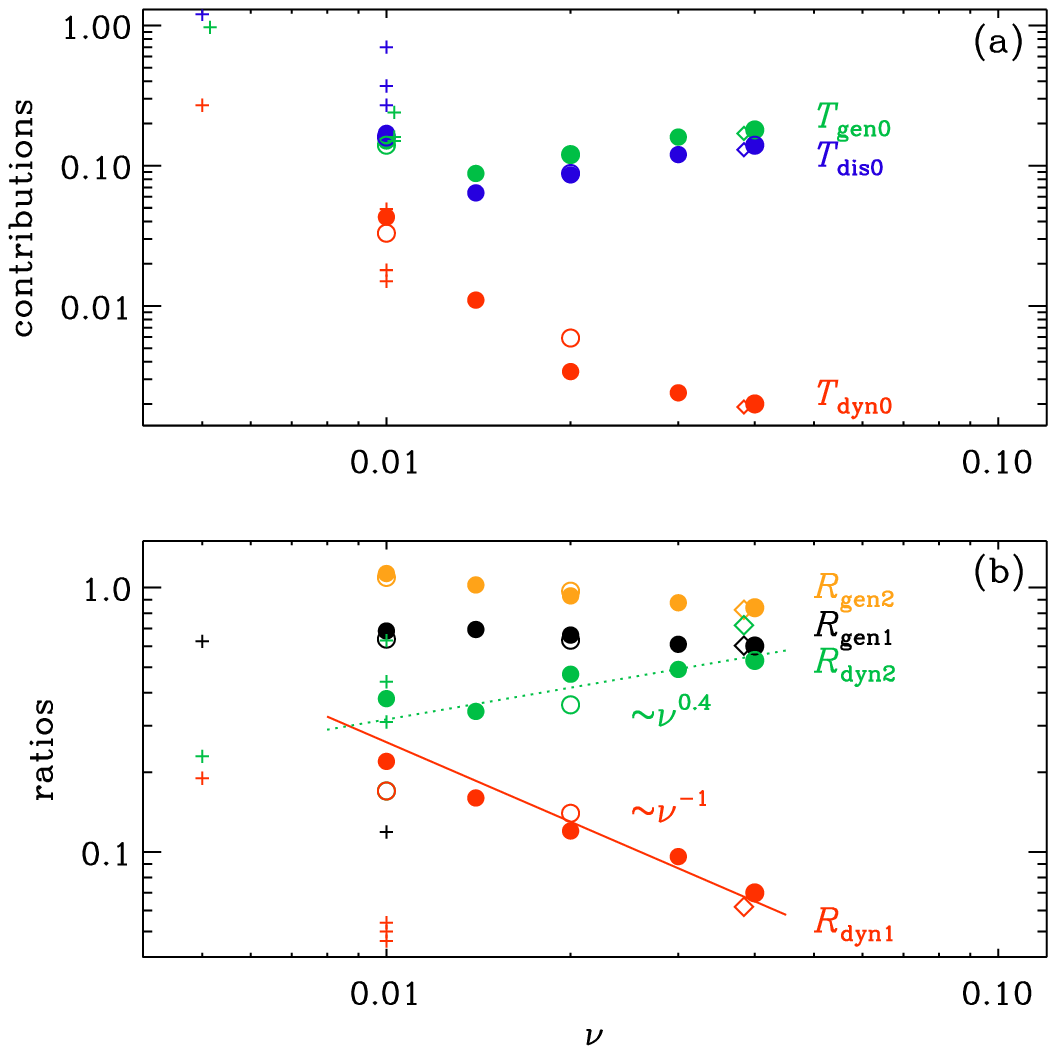}
\end{center}\caption{
Scaling of (a) the ratios $R_\mathrm{gen1}$, $R_\mathrm{gen2}$, $R_\mathrm{dyn1}$, and $R_\mathrm{dyn2}$,
and (b) the early peak values of the terms $T_\mathrm{gen0}$, $T_\mathrm{dis0}$, and $T_\mathrm{dyn0}$ with $\nu$.
Runs~C and F are lower-resolution results of Runs~B and E, respectively, and are shown as open symbols.
Run~H is a lower-resolution result of Runs~I and is shown as a diamond.
Runs~A, N, O, and P are results obtained with upwinding and are shown as plus signs.
The red line is proportional to $\nu^{-1}$, which is the approximate scaling found for $R_\mathrm{dyn}$.
The approximate $\nu^{+0.4}$ scaling for $R_\mathrm{dyn2}$ is more uncertain.
}\label{pres}\end{figure}

Given that the vorticity generation is proportional to $\nu$,
one might expect the amount of vorticity production to be proportional to the value of $\nu$.
To analyze this, we have chosen two representative values of $R_\mathrm{gen}$; see \Fig{pom2_Hy1024c_noshock}(b),
where we have marked as a red line the average value of the local plateau of $R_\mathrm{gen}$ during the collapse phase ($1.3\leq t\leq2.7$),
as well as the early maximum at $t\approx0.6$.
We refer to these two values as $R_\mathrm{gen1}$ and $R_\mathrm{gen2}$, respectively.
In a few cases, however, no local plateau or local maximum of $R_\mathrm{gen}$ exists.
Those cases are marked in \Tab{TSummary} with a dash.
In \Fig{pres}, we plot the $\nu$ dependence of $R_\mathrm{gen1}$ and $R_\mathrm{gen2}$,
along with two representative values of the dynamo-like term, $R_\mathrm{dyn1}$ and $R_\mathrm{dyn2}$,
taken as an average during the early phase ($0.5\leq t\leq1.3$) and the peak shortly before the end of the collapse.
Finally, we plot the early peak values (before $t=0.5$) of the terms $T_\mathrm{gen0}$, $T_\mathrm{dis0}$, and $T_\mathrm{dyn0}$.

%FIG5

We clearly see that neither $R_\mathrm{gen1}$ nor $R_\mathrm{gen2}$ are proportional to the value of $\nu$, as one might have naively expected.
Both ratios slightly increase toward smaller values of $\nu$, but only $R_\mathrm{gen2}$ exceeds unity,
so vorticity generation is only expected at the early phase, before the actual collapse.
We also see that both $T_\mathrm{dyn}$ and $R_\mathrm{dyn}$ increase toward smaller values of $\nu$.
In particular, we find that $R_\mathrm{dyn1}\propto\nu^{-1}$; see \Fig{pres}.
For $R_\mathrm{dyn2}$, the trend is more uncertain, but possibly compatible with $\nu^{+0.4}$, i.e., the other way around.

Although $R_\mathrm{dyn}$ is still small compared to unity, it raises the question whether self-amplification might be possible for sufficiently small viscosities.
Indeed, there has been a lot of work regarding the possibility of dynamo-like vorticity generation \citep{Krause+Ruediger74, Moiseev+83, Levina19}.
We know, however, that the closest analogy to the $\alpha$ effect in mean-field electrodynamics is the anisotropic kinetic $\alpha$ effect,
also known as the AKA effect \citep{Frisch+87, Sulem+89}, which is represented by a rank-3 tensor.
It was found that such systems require progressively larger scale separation as the Reynolds number increases;
see \cite{BvR01}, who found that above a Reynolds number of about eight, no large-scale flow production occurred.
Large-scale vorticity production is also possible in the presence of shear \citep{Elperin+03,Kapyla+09}.
To address the question of small-scale vortical flow production by the dynamo-like term $T_\mathrm{dyn}$,
we would need to use much larger resolution than what has currently been possible.
Note also that for $\nu=0.01$, the higher resolution result ($2048^3$ instead of $1024^3$) shows
slightly larger values for $R_\mathrm{dyn}$ and $T_\mathrm{dyn}$.
Thus, poor resolution might underestimate the dynamo-like term.

Whether or not vorticity amplification through the $R_\mathrm{dyn}$ term is possible is still unclear.
If the current trend were to continue toward smaller viscosities, it would exceed unity for $\nu\approx10^{-3}$, suggesting the possibility of a vorticity dynamo.
This value of $\nu$ would be five times larger than that of Run~A.
Looking at \Tab{TSummary}, it would therefore correspond to $\Rey\approx400$.
Note, however, that this amplification would occur early in the run and is not related to the actual collapse.
Near the end of the collapse, there is a second amplification phase, but it tends to peak well in the range
where the results begin to be unreliable.
Furthermore, looking at \Fig{pres}, we see that the $\nu$ dependence of $R_\mathrm{dyn2}$ is possibly
the other way around, i.e., amplification becomes weaker for smaller viscosities.
On the other hand, this result itself may well be flawed and unreliable.
In any case, studying the possibility of a vorticity dynamo requires a more dedicated approach that is beyond
the present scope, where we focus on collapse-driven vorticity amplification.

\subsection{Velocity, density, and vorticity spectra}
\label{VelocitySpectra}

\begin{figure}\begin{center}
\includegraphics[width=\columnwidth]{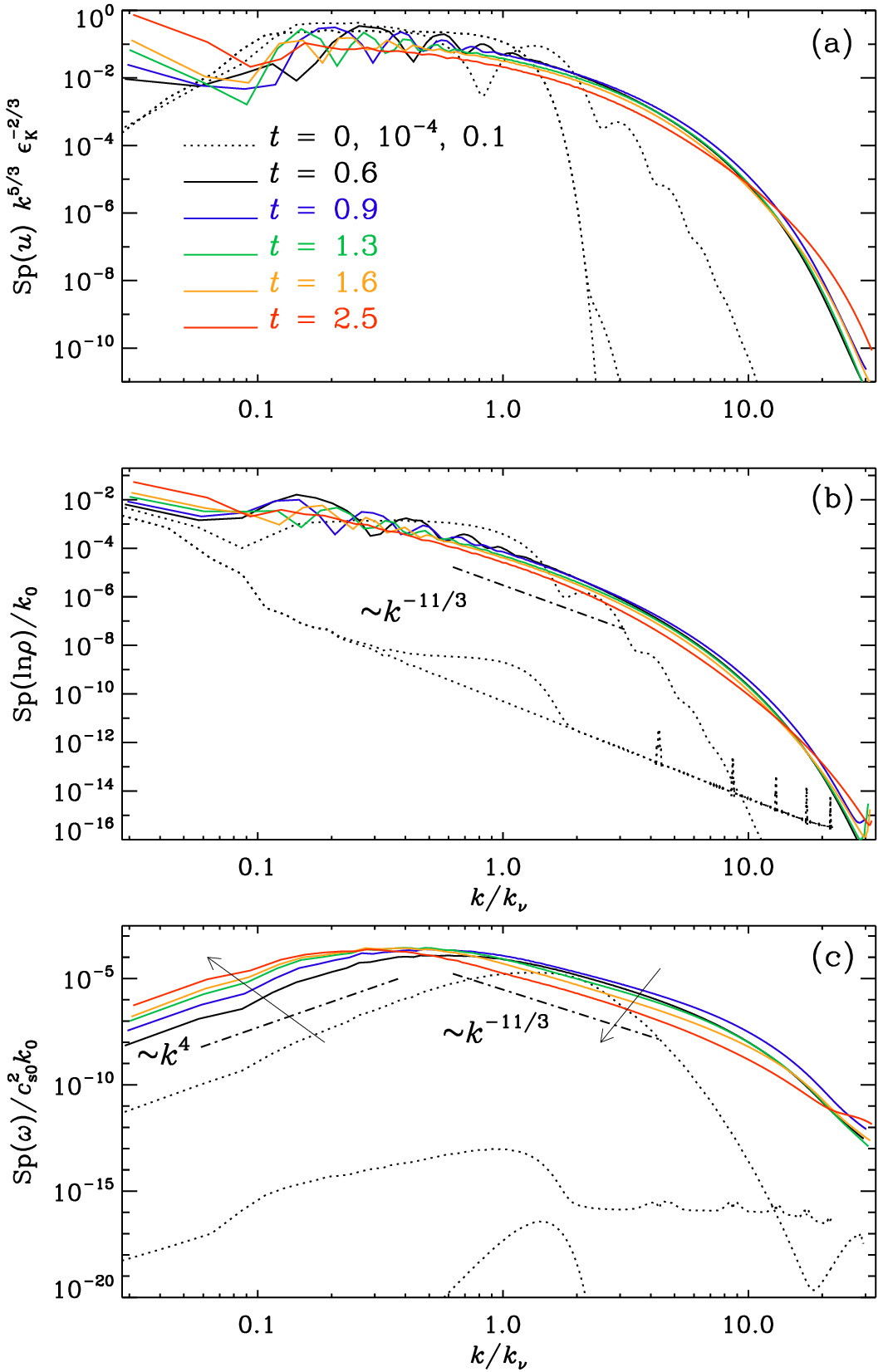}
\end{center}\caption{
(a) Compensated spectra of $\uu$, and uncompensated spectra of (b) $\ln\rho$ and (c) $\oo$,
for Run~B with $\nu=0.01$ at $t=0$, $10^{-4}$, and 0.1 (all dotted lines),
as well as 0.6 (solid black), 0.9 (blue), 1.3 (green), 1.6 (orange), and 2.5 (red).
Note the $k^4$ subinertial range spectrum in $\oo$, in analogy to a similar behavior of the magnetic field in MHD.
The approximate $k^{-11/3}$ spectrum is unrelated to a Kolmogorov spectrum
and appears here in the beginning of the diffusive subrange.
}\label{pspec_Hy2048a_noshock}\end{figure}

Let us now inspect the scale dependence of velocity, density, and vorticity.
For that purpose, we plot spectra of $\uu$, $\ln\rho$, and $\oo$ at different times;
see \Fig{pspec_Hy2048a_noshock}.
Here, we have compensated the velocity spectrum by $\epsK^{-2/3}k^{5/3}$
and the vorticity spectrum is left uncompensated, but normalized with $\csz^2 k_0$.

%FIG6

In spite of the rather sharp and early exponential cutoff of the initial velocity with $k_\mathrm{cut}=25$
(corresponding to $k_\mathrm{cut}/k_\nu=1.7$), a fairly long dissipative subrange develops.
It appears much longer than what one usually sees in nearly incompressible forced turbulence \citep[see, e.g.,][]{BRS23}.
There is no ``bottleneck'', i.e., no spectral bump just before the dissipative subrange \citep{Falk94}.
The actual inertial range is also a bit steeper than a usual Kolmogorov spectrum.
This is expected when the flow is dominated by shocks \citep{KP73}.
The oscillations in the spectrum at early times travel to the left.
They are a standard phenomenon that is caused by having initialized a velocity field without
adding corresponding perturbations in the density \citep{BN22, Sharma+23}.
At low wavenumbers, we see the development of the Jeans instability with a spectral peak at the box wavenumber,
similarly to what has been seen in earlier work; see Figure~3 of \cite{BN22} and Figure~4 of \cite{Schober2026}.

%FIG7
\begin{figure*}\begin{center}
\includegraphics[width=\textwidth]{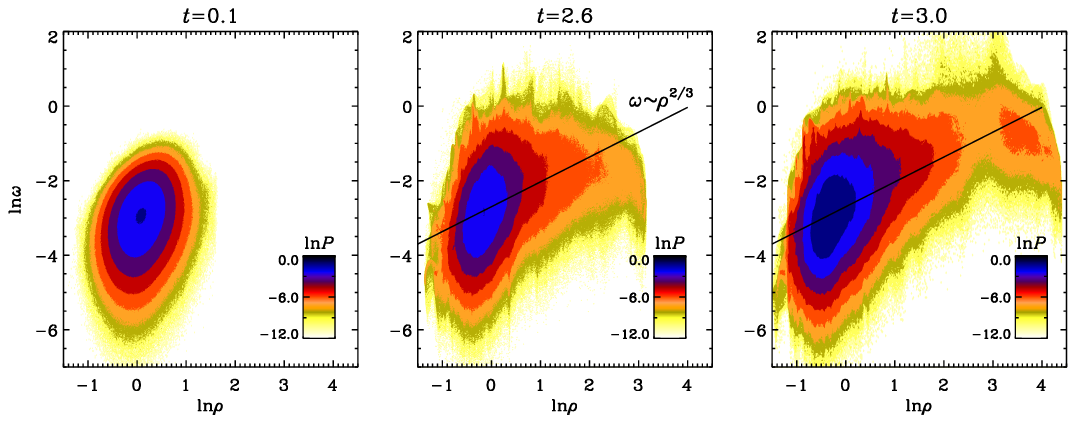}
\end{center}\caption{
Two-dimensional probability density functions $P(\ln\omega,\ln\rho)$ near the end of Run~N.
}\label{ppdf2d_comp_BE_3times}\end{figure*}

Spectra of logarithmic density were previously found to be good proxies of the
irrotational part of the velocity spectrum \citep{BS25}.
They are roughly similar to the total kinetic energy spectra.
This is because of the strong dominance of the irrotational flow component.
The small spikes in $\Sp(\ln\rho)$ at the earliest times occur at $k=69$ and higher multiples,
and are presumably a consequence of having truncated the Bonnor--Ebert sphere at $|\xx|=2\pi$.
The actual spectra of the irrotational part of the velocity are shown in \App{IrrotationalVelocitySpectra}.

In the vorticity spectrum, we see a rapid build-up of a $k^4$ subinertial range.
In the turbulent inertial range ($0.1\leq k/k_\nu\leq1$), the vorticity spectrum is flat,
but followed by an approximate $k^{-11/3}$ subrange.
During the developed collapse phase, the spectral vorticity decays at high wavenumbers,
but there is a growth at low wavenumbers, akin to the inverse cascade found previously for the magnetic field.
This finding suggests some analogy between vorticity and the magnetic field in MHD,
but it is here just a direct consequence of the Jeans instability driving random fluctuations in the velocity.
A random velocity spectrum would be proportional to $k^2$ and would lead to a $k^4$ vorticity spectrum.
We do not really see a $k^2$ velocity spectrum at low $k$.
This is presumably because of a strong dominance of the radial inflow.
However, we have not tried to separate the mean from the fluctuating flows.

\subsection{Vorticity concentration during collapse}

\FFig{pVAR} suggested that preexisting vorticity just gets more concentrated toward the center as the collapse proceeds.
To examine this in more detail, we show in \Fig{ppdf2d_comp_BE_3times} two-dimensional probability density functions $P(\ln\omega,\ln\rho)$ near the end of Run~N.
Similar probability density functions are routinely examined for the magnetic field $\BB$.
In that case, one finds a correlation $|\BB|\propto\rho^{2/3}$ because of magnetic flux conservation \citep{Crutcher2012,BN23}.
In the present case, we see that, toward the end of the collapse, the vorticity begins to show an additional hump in the extension of the line $\omega\propto\rho^{2/3}$.
This is analogous to the magnetic field vs.\ density correlation, provided the mean circulation, $\oint\uu\cdot\dd\bm{\ell}=\int\oo\cdot\dd\SSS$, is conserved.
This is the case here if the viscosity is negligible.

\subsection{Does the collapse produce additional vorticity?}

\begin{figure}\begin{center}
%dir1='Hy512d_noshock'         & nu1=2e-2 & ampl1=.2
%dir2='Hy512d_noshock_ampl01'  & nu2=2e-2 & ampl2=.1
%dir3='Hy512d_noshock_ampl005' & nu3=2e-2 & ampl3=.05
\includegraphics[width=\columnwidth]{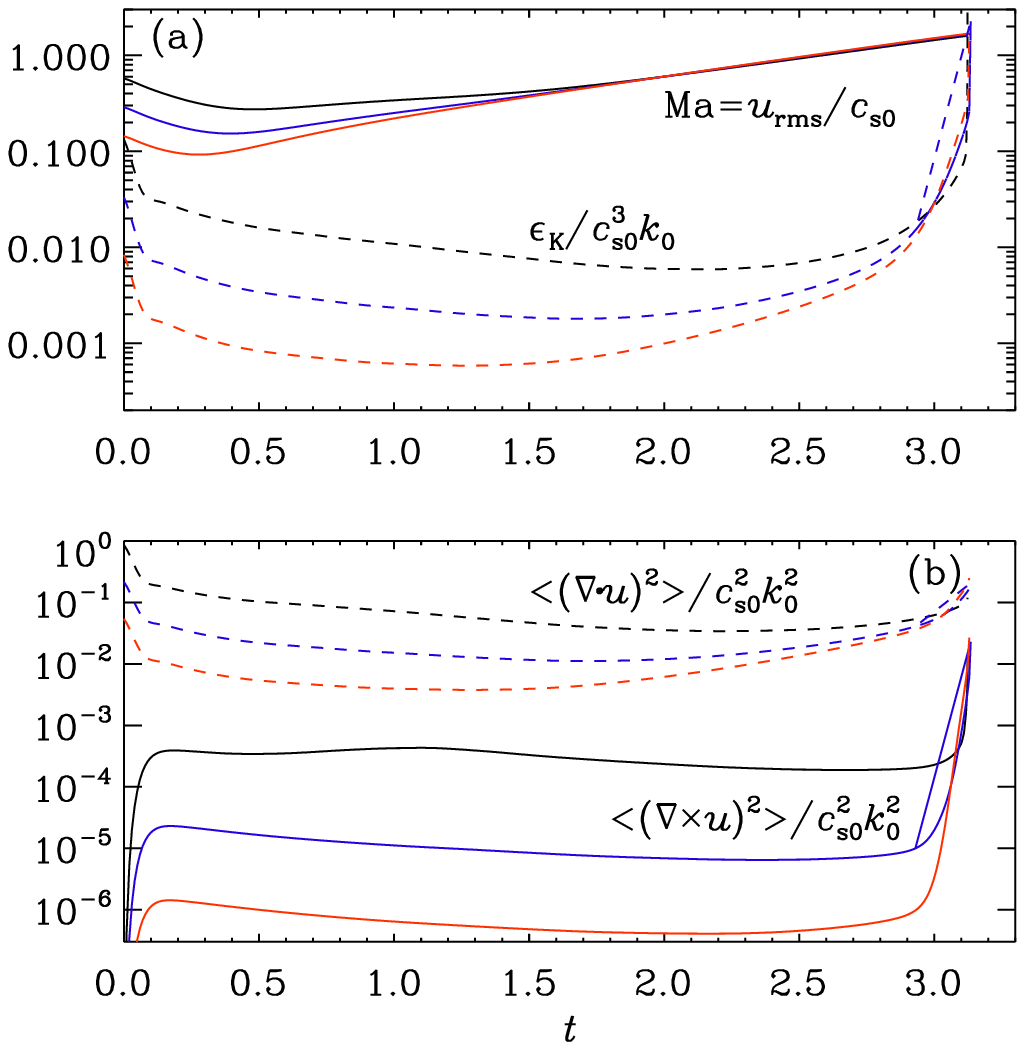}
\end{center}\caption{
Similar to \Fig{pcomp_ts}, but for different flow amplitudes:
$u_\mathrm{ini}=0.2$ (red line), $0.1$ (black line), and $0.05$ (blue line),
corresponding to Runs~F, J, and K.
}\label{pcomp_ts_ampl}\end{figure}

If the collapse were to produce additional vorticity (on top of the vorticity produced by the viscous conversion from the initially irrotational flow),
the resulting vorticity should not strongly depend on the initial level of turbulence.
In \Fig{pcomp_ts_ampl}, we present a plot similar to \Fig{pcomp_ts},
but now we compare the Mach number, the nondimensional kinetic energy dissipation, and
the mean square velocity divergence and vorticity for different initial flow amplitudes.

%FIG8

\begin{figure}\begin{center}
\includegraphics[width=\columnwidth]{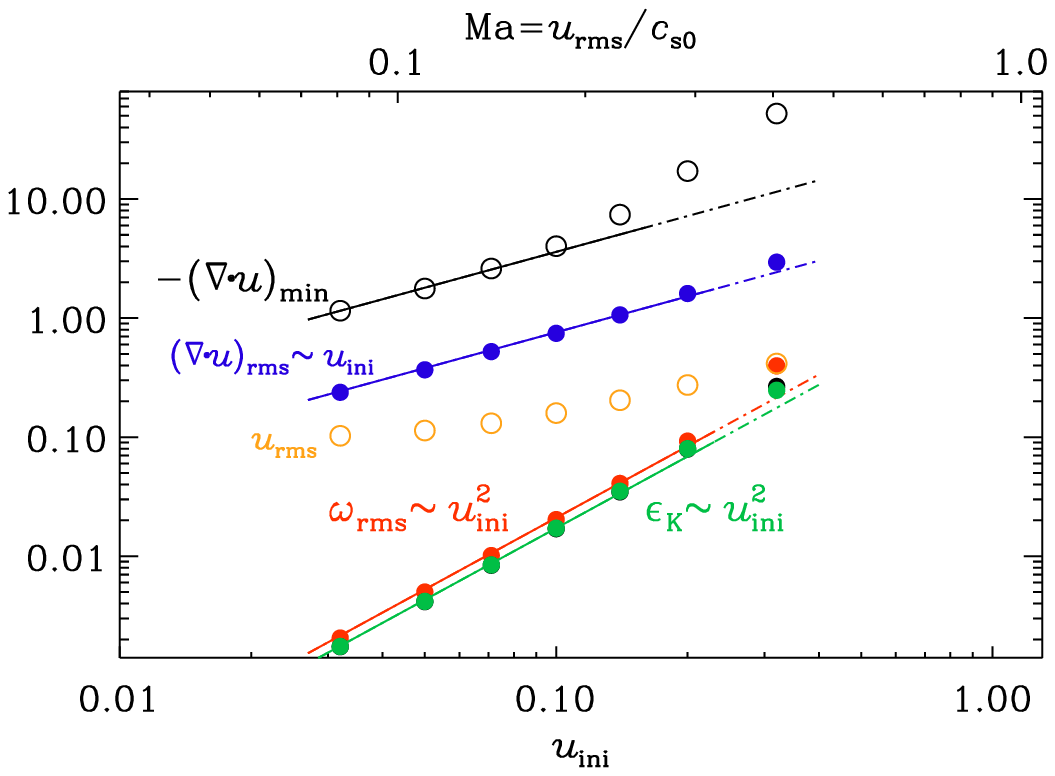}
\end{center}\caption{
Scalings of $(\nab\cdot\uu)_\mathrm{rms}$, $\orms$, and $\epsK$ with $u_\mathrm{ini}$ at $t=0.5$.
The data points include Runs~F, J, and K.
The open black symbols depict $-(\nab\cdot\uu)_\mathrm{min}$.
They clearly deviate from a power-law scaling.
The upper horizontal axis gives the Mach number based on the rms velocity at $t=0.5$,
which is also shown as open orange symbols.
They also deviate from a power-law scaling.
}\label{ptab_ampl}\end{figure}

It turns out that, while the resulting flow amplitude, as measured by
the Mach number, is indeed independent of the initial flow altitude, all other turbulence-specific quantities are not.
In particular, the rms flow divergence, $(\nab\cdot\uu)_\mathrm{rms}$ is found to be proportional to $u_\mathrm{ini}$,
while $\orms$ is proportional to $u_\mathrm{ini}^2$.
This is shown in \Fig{ptab_ampl}, where we used the values of $(\nab\cdot\uu)_\mathrm{rms}$ and $\orms$ with $u_\mathrm{ini}$ at a relatively early time $t=0.5$,
when this scaling is not yet much affected by the collapse.
At later times, the scaling of $\orms$ is still similar, but that of $(\nab\cdot\uu)_\mathrm{rms}$ becomes shallower due to the collapse.
Since the separation between the three curves for $\orms$ remains unchanged until $t=2.7$,
it strongly suggests that the gravitational collapse has no direct effect on the vorticity production.
We argue that vorticity production is rather a secondary effect resulting from the amplification of initial flow divergences,
which then produce vorticity through the action of viscosity.

%FIG9

In \Fig{ptab_ampl}, we also show the $u_\mathrm{ini}$ dependence of the negative minimum velocity divergence,
$-(\nab\cdot\uu)_\mathrm{min}$.
For $\Ma>0.2$, it deviates from a power-law scaling and becomes steeper for larger values of $u_\mathrm{ini}$.
It is tempting to associate the term $-(\nab\cdot\uu)_\mathrm{min}$ with a measure of the occurrence of shocks,
which we expect to happen for Mach numbers around unity.
Since the dependence of $\orms$ on $u_\mathrm{ini}$ follows a power law for all values of $u_\mathrm{ini}$,
and not just for $u_\mathrm{ini}\ga0.1$, as seen from the open symbols in \Fig{ptab_ampl},
we can conclude that vorticity production is not solely a consequence of shocks.

\subsection{Comparison with noncollapsing turbulence}
\label{NoncollapsingTurbulence}

\begin{figure}\begin{center}
%dir1='Hy512h_noshock2'    & nu1=1e-2 & k0_1=5.  ;(black)
%dir2='Hy512h_noshock_k10' & nu2=1e-2 & k0_2=10. ;(blue)
%dir3='Hy512h_noshock_k20' & nu3=1e-2 & k0_3=20. ;(red)
\includegraphics[width=\columnwidth]{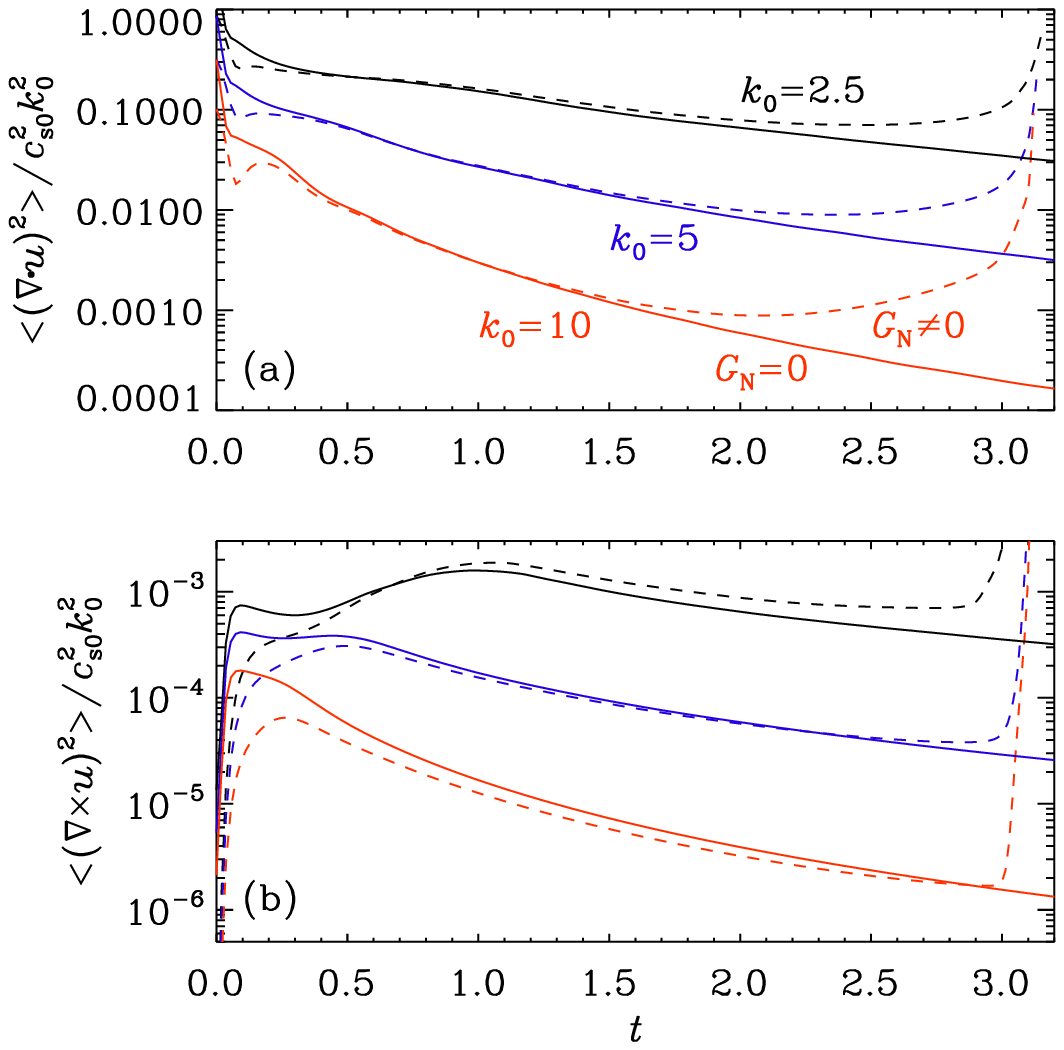}
\end{center}\caption{
Similar to \Fig{pcomp_ts_ampl}, but for different initial wavenumbers:
$k_0=2.5$ (black line), $5$ (blue line), and $10$ (red line), corresponding to Runs~N, O, and P.
Dashed lines show results from runs with self-gravity ($G_\mathrm{N}\neq 0$) and 
solid lines for runs without self-gravity ($G_\mathrm{N}=0$).
}\label{pcomp_ts_k0}\end{figure}

To compare our results with noncollapsing turbulence, we show in \Fig{pcomp_ts_k0} the time dependence of the mean square velocity divergence and vorticity
with and without gravity for three values of $k_0$.
Especially for large values of $k_0$ (5 and 10), we clearly see that the interval of developed collapse (see \Sec{sec:Summary_runs} for our definition of developed collapse) is characterized by
an enhanced production of flow divergences, but the vorticity remains at the original level that is also obtained without collapse ($G_\mathrm{N}=0$).
This shows once again that vorticity emerges through viscous conversion and is weak if the level of irrotational turbulence is weak.
For larger eddies of the initial turbulence (smaller values of $k_0$), the absolute level of irrotational turbulence is larger,
and therefore the resulting vorticity is also larger at early times.
In all cases, however, a sharp increase in the vorticity is only seen at $t=3$, which is when the simulation becomes underresolved.

%FIG10

Given that the vorticity is just inherited from the initial turbulence,
one might have expected that $k_\omega=\orms/\urms$, and therefore also $\orms$ itself, increase with $k_0$,
but \Fig{pcomp_ts_k0} shows that the opposite is the case and that the lines with larger values of $k_0$ are below those with smaller values of $k_0$.
This has to do with the fact that both $\drms$ and $\orms$ are strongly characterized by the value of $k_\mathrm{cut}$ in \Eq{kcut}.
Increasing the value of $k_0$ therefore lowers the total integrated flow divergence, which causes the decline.
This aspect goes beyond the interest of the present paper.
Therefore, we refer to \App{k0Dependence} for a plot of the $k_0$ dependence of $\drms$ and $\orms$.

%FIG11
\begin{figure}\begin{center}
\includegraphics[width=\columnwidth]{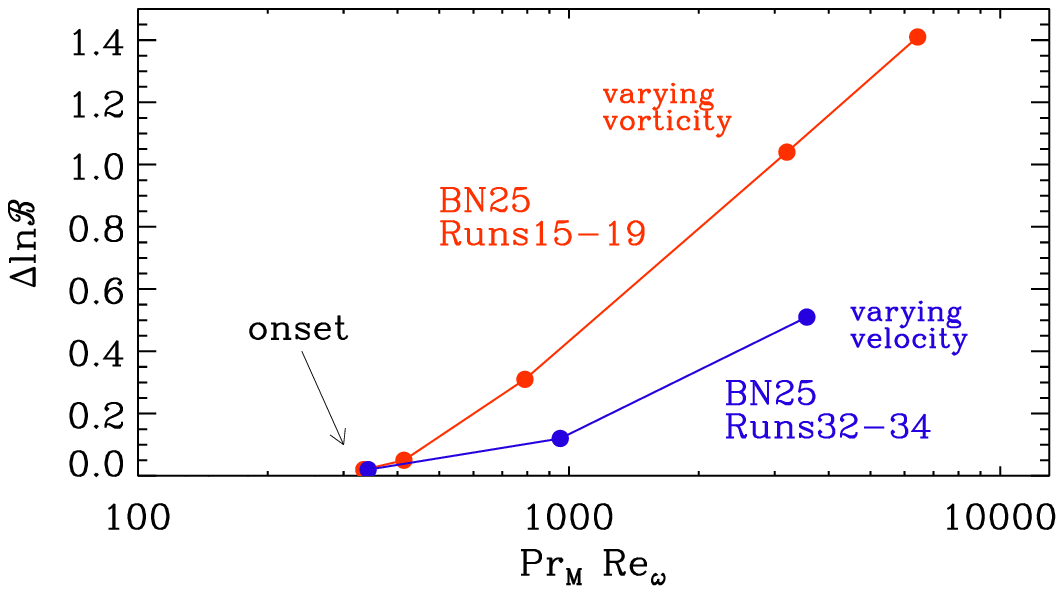}
\end{center}\caption{
Dependence of $\Delta\ln\mathcal{B}$ vs.\ $\Pm\Rey_\omega$ for Runs~15--19 and 32--34 from \cite{BN25},
showing the critical onset value of $\Pm\Rey_\omega$ being around 300.
}\label{ptab_BN25}\end{figure}

\subsection{Critical vorticity for dynamo action}

The question regarding the critical value of the vorticity for small-scale dynamo action has already been addressed by \cite{BN25}.
Here, we revisit this question by plotting in \Fig{ptab_BN25} the magnetic field amplification for two families of runs.
The basic idea is that the onset of small-scale dynamo action may be determined by the value of the magnetic vorticity Reynolds number, i.e.,
\begin{equation}
\Pm\Rey_\omega=\orms/\eta k_0^2,
\label{PmReyOm}
\end{equation}
where $\Pm=\nu/\eta$ is the magnetic Prandtl number and $\Rey_\omega=\orms/\nu k_0^2$ is the vorticity Reynolds number \citep{HBM04,Elias-Lopez+23, Elias-Lopez+24}.
In \Fig{ptab_BN25}, we plot the logarithmic growth, $\Delta\ln\mathcal{B}$, of the magnetic field normalized by the instantaneous equipartition field strength, $\mathcal{B}$,
against $\Pm\Rey_\omega$ for the Runs~15--19 and 32--34 of \cite{BN25}.
In the first family of runs, the magnetic Reynolds number was kept approximately constant, and only the degree of irrotationality was increased, i.e., the vorticity was decreased,
while in the second family of runs, the irrotationality was kept constant and the magnetic Reynolds number was increased, i.e., the velocity was increased.
We clearly see a consistent bifurcation away from zero for $\Pm\Rey_\omega\approx300$.
This reinforces their earlier claim that the onset of small-scale dynamo action is determined by the magnetic vorticity Reynolds number.
If correct, it would also mean that purely irrotational flows would not act as dynamos, contrary to some earlier theoretical works \citep{Kazantsev+85, Afonso+19}.
However, although we did not consider magnetic fields, this would not affect our current conclusion that the collapse itself would not power small-scale dynamos.

\section{Conclusions}

Following earlier recommendations \citep{BN22,BN25}, it is worth revisiting earlier claims of turbulence generation and the resulting dynamo action.
Here, we have paid particular attention to the importance of direct numerical simulations.
It is possible that certain subgrid-scale schemes could lead to excessive vorticity generation.
In any case, in the current model of barotropic turbulence, which has been studied numerically in many previous studies
\citep[see also][]{Elias-Lopez+23, Elias-Lopez+24},
vorticity production from a strictly irrotational flow is only possible through the action of viscosity.
This should also be possible in ideal codes, where total energy, mass, and momentum conservation are enforced, which implies entropy production from shocks;
see \cite{Federrath+11, Federrath+11b} and \cite{Porter+15}, for example.
However, note that the resulting vorticity production is a continuous function of the Mach number.
For example, \cite{BN25} found that vorticity production is proportional to $\Ma^{1.6}$ and $\Ma_0^{0.84}$, where $\Ma$ and $\Ma_0$ are the actual and initial Mach numbers.
By contrast, the values of $-(\nab\cdot\uu)_\mathrm{min}$, which we have taken as a measure of the
occurrence of shocks, deviate from a power-law dependence on $u_\mathrm{ini}$ (or the Mach number).
We have therefore argued that vorticity production is not solely a consequence of shocks.

Following earlier work of \cite{MB06}, we have also revisited the role of the vortex dynamo-like term $T_\mathrm{dyn}$ in \Eq{Tdyn_appearance}.
It turns out that it grows $\propto\nu^{-1}$; see the red data points in \Fig{pres}.
Again, this result is not collapse-related, but it would be useful to study the possibility of vortex dynamo-like amplification,
which might occur for Reynolds numbers above about 400.

To address the question of the extent to which gravitational collapse plays a role in the generation of vorticity,
we have varied the initial irrotational flow amplitude, $u_\mathrm{ini}$.
It turned out that, while it had no noticeable effect on the rms velocity,
it had a direct effect on the production of velocity divergences with $(\nab\cdot\uu)_\mathrm{rms}\propto u_\mathrm{ini}$,
and an even stronger effect on the production of vorticity with $\orms\propto u_\mathrm{ini}^2$.
This strongly suggests that, in our simulations, the gravitational collapse has no effect on vorticity production 
and that it is rather a secondary effect from the viscous conversion of initial flow divergences.
Thus, our work suggests once again that vorticity generation from a barotropic gravitational collapse 
is not easily obtained---even at the fairly high numerical resolutions available today.

Future work using adaptive mesh refinement (AMR) might help clarify the role of the collapse in producing vorticity.
It should be kept in mind, however, that AMR might introduce numerical noise at the grid level.
This could affect the results, especially if a strongly violent collapse triggers more than one refinement level between two time steps.
An alternative might be the transformation to a collapsing coordinate system \citep{Robertson+Goldreich12,Murray+17,BN25,Irshad+26}.
This would most adequately address the concern about noise at the grid level.
The goal would be to show that the rms vorticity over the full domain---and not just the collapsing subdomain---increases due to the collapse
and not just due to the viscous conversion of the originally irrotational turbulence.
Once this is demonstrated, and once the resulting magnetic vorticity Reynolds number, as defined in \Eq{PmReyOm},
exceeds the critical onset value of 300, we can begin to rely on collapse-induced small-scale dynamo action.

\begin{acknowledgements}
This work emerged during the Focus Week on ``Magnetic Fields in High-Redshift Galaxies'' at the Institute for Fundamental Physics of the Universe in Trieste.
We thank the organizers for having provided such a stimulating atmosphere.
We also thank Irshad P, Kandaswamy Subramanian, Sharanya Sur, Enrique V{\'a}zquez-Semadeni,
and the anonymous referee for useful comments on the manuscript.
This research was supported in part by the European Research Council through the ERC Synergy Grant COSMOMAG under grant No.\ 101224803,
the Swedish Research Council (Vetenskapsr{\aa}det) under grant No.\ 2025-05957,
the National Science Foundation under grant Nos.\ NSF AST-2307698, AST-2408411, and NASA Award 80NSSC22K0825.
We acknowledge the allocation of computing resources provided by the
Swedish National Allocations Committee at the Center for Parallel Computers at the Royal Institute of Technology in Stockholm.
EN acknowledges funding from the Italian Ministry for Universities and Research (MUR) through the ``Young Researchers'' funding call (Project MSCA 000074).

\vspace{2mm}\noindent
{\em Software and Data Availability.} The source code used for
the simulations of this study, the {\sc Pencil Code} \citep{PC},
is freely available on \url{https://github.com/pencil-code/} with its latest developments.
The simulation setups and corresponding data are freely available on
\href{https://doi.org/10.5281/zenodo.21109846}{DOI:10.5281/zenodo.21109846}, as well as on
\href{https://norlx65.nordita.org/~brandenb/projects/BEcollapse/}{norlx65.nordita.org/~brandenb/projects/BEcollapse}.
\end{acknowledgements}

\appendix
\section{Different density profiles}
\label{DifferentDensityProfiles}

\begin{figure}\begin{center}
\includegraphics[width=\columnwidth]{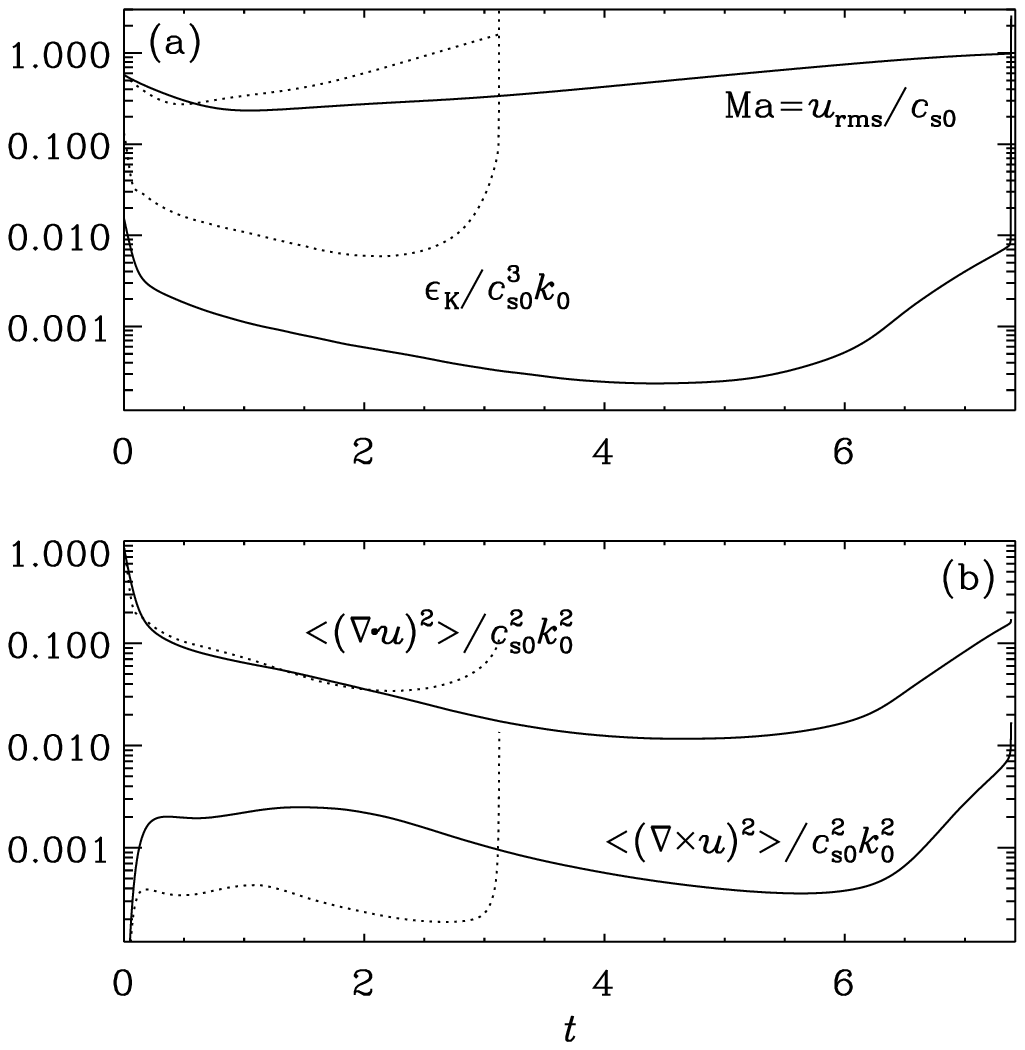}
\end{center}\caption{
Similar to \Fig{pcomp_ts}, but for $\rho=\exp(-\psi)$ (solid line) instead of the $\rho=\exp[\exp(-\psi)]$ profile (dotted line).
}\label{pcomp_ts_nopretend}\end{figure}

We have emphasized in \Sec{sec:Summary_runs} that the results for different density profiles are rather small.
To demonstrate this, we present in \Fig{pcomp_ts_nopretend} the evolution of the Mach number, the nondimensional kinetic energy dissipation,
and the mean square velocity divergence and vorticity for different initial flow amplitudes for Run~Q with
$\rho=\exp(-\psi)$ instead of the original $\rho=\exp[\exp(-\psi)]$ profile.
It turns out that the collapse time is longer, but the evolution is qualitatively similar to that with the double exponential profile.

%FIG12

\section{Irrotational velocity spectra}
\label{IrrotationalVelocitySpectra}

\begin{figure}\begin{center}
\includegraphics[width=\columnwidth]{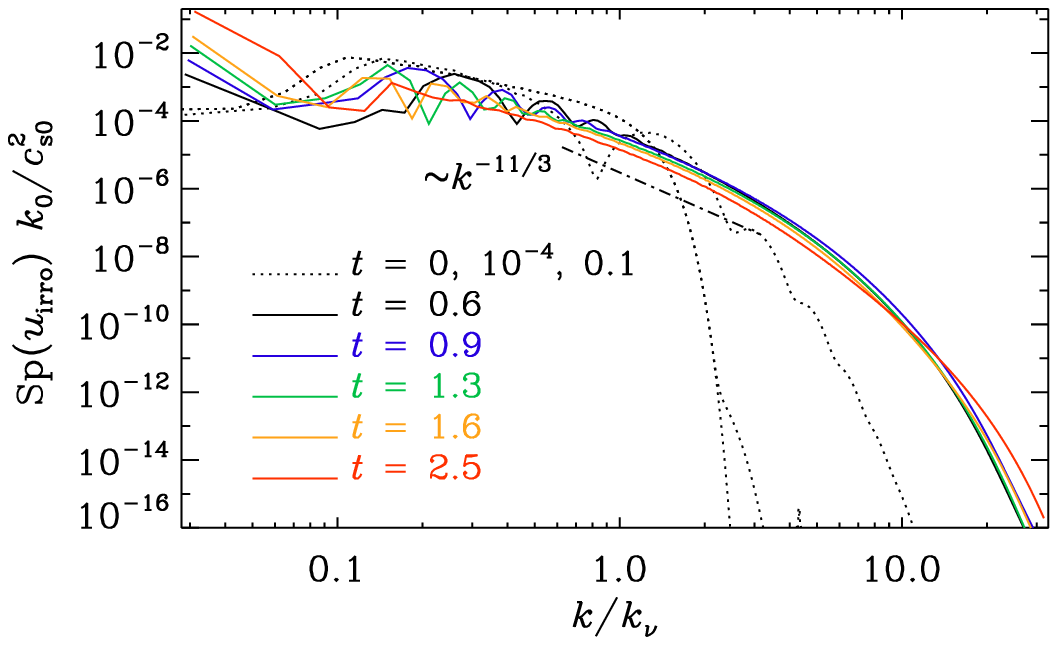}
\end{center}\caption{
Uncompensated spectra of $\uu_\mathrm{irro}$
for Run~B with $\nu=0.01$ at $t=0$, $10^{-4}$, and 0.1 (all dotted lines),
as well as 0.6 (solid black), 0.9 (blue), 1.3 (green), 1.6 (orange), and 2.5 (red).
}\label{pspec_irro_Hy2048a_noshock}\end{figure}

In \Sec{VelocitySpectra}, we presented spectra of the logarithmic density as proxies of the irrotational part of the velocity
and referred to \cite{BS25} for earlier work demonstrating this in magnetically modified turbulence.
In \Fig{pspec_irro_Hy2048a_noshock}, we present spectra of the actual irrotational part of the velocity for Run~B.
As in \cite{BS25}, we compute $\Sp(\uu_\mathrm{irro})$ as the difference between $\Sp(\uu)$ and $\Sp(\oo)/k^2$.
The spectra of the irrotational part of the velocity are indeed similar to the logarithmic density spectra, except that they do not show 
the small spikes that we saw in $\Sp(\ln\rho)$ at the earliest times.

%FIG13

%FIG14
\begin{figure}\begin{center}
\includegraphics[width=\columnwidth]{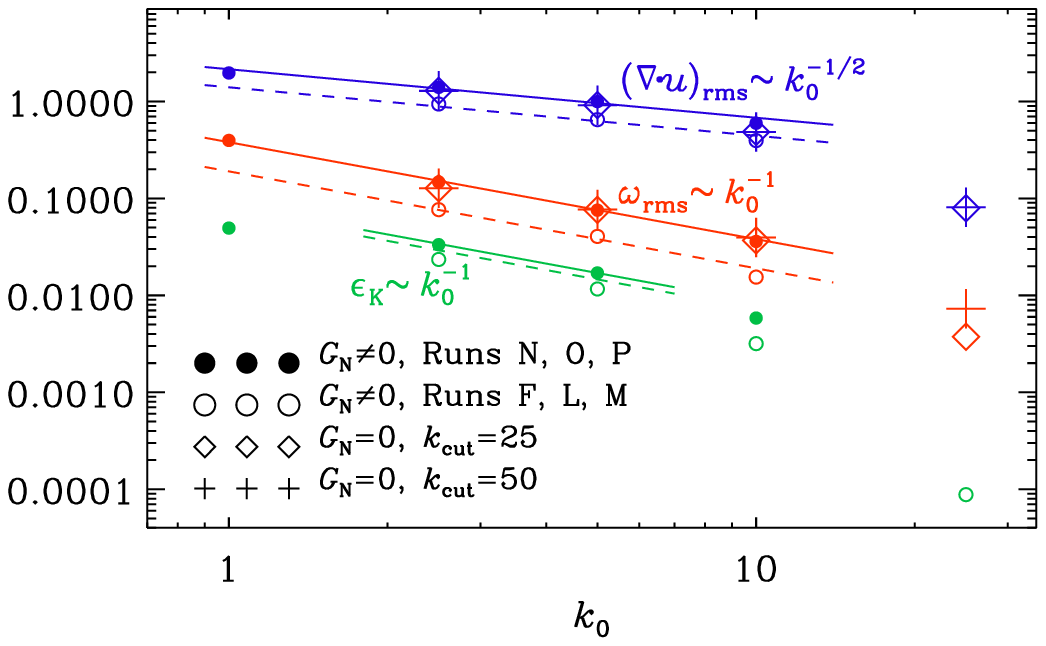}
\end{center}\caption{
Scalings of $\drms\equiv(\nab\cdot\uu)_\mathrm{rms}$ and $\orms$ with $k_0$ at $t=2.0$.
}\label{ptab_k0}\end{figure}

\section{Dependence on $k_0$}
\label{k0Dependence}

In \Sec{NoncollapsingTurbulence}, we presented results for different values of $k_0$ and found that
$\drms$, $\orms$, and $\epsK$ decrease with increasing values of $k_0$.
\FFig{ptab_k0} shows these values at $t=2$ as a function of $k_0$.
It turns out that, in the limited parameter range investigated, the $k_0$ dependences are well approximated by
$\drms\propto k_0^{-1/2}$, $\orms\propto k_0^{-1/2}$, and also $\epsK\propto k_0^{-1/2}$ 
We emphasize, however, that this result is independent of the collapse dynamics.
This is clearly seen by comparing with the $G_\mathrm{N}=0$ versions of Runs~N, O, and P.
It also turns out that, except for large values of $k_0$, the values of $\drms$, $\orms$, and $\epsK$ are not very sensitive to the value of $k_\mathrm{cut}$.

\bibliographystyle{aasjournal}
\bibliography{ref}

\end{document}